\documentstyle[aps,preprint,tighten,eqsecnum,epsf,floats]{revtex}
\def\d{\partial}
\def\bfnab{\mbox{\boldmath$\nabla$}\!}
\def\+{\dagger}
\def\x{{\mbox{\boldmath $x$}}}
\def\j{{\mbox{\boldmath $j$}}}
\def\v{{\mbox{\boldmath $v$}}}
\def\z{{\mbox{\boldmath $z$}}}
\def\phiA{\varphi_{\rm A}}
\def\phiB{\varphi_{\rm B}}
\def\A{{\rm A}}
\def\B{{\rm B}}
\begin{document}

\title{Domain walls of relative phase in two-component Bose-Einstein
condensates} 
\author{D.T.~Son$^{1,3}$ and M.A.~Stephanov$^{2,3}$}
\address{$^1$ Physics Department, Columbia University, New York, 
New York 10027}
\address{$^2$ Department of Physics, University of Illinois, Chicago, 
Illinois 60607-7059}
\address{$^3$ RIKEN-BNL Research Center, Brookhaven National Laboratory,
Upton, New York 11973}

\maketitle

\begin{abstract}
We consider a system of two interpenetrating Bose-Einstein condensates
of atoms in two different hyperfine spin states.  We show that in the presence
of a small coupling drive between the two spin levels, there exist
domain walls across which the relative phase of the two condensates
changes by $2\pi$.  We give the physical interpretation of such walls.
We show that the wall tension determines the force between certain
pairs of vortices at large distances.  We also show that the probability of
the spontaneous decay of the domain wall is exponentially suppressed,
 both at finite and at zero temperature, and
determine the exponents in the regime of small Rabi frequency.  We
briefly discuss how such domain walls could be created in
future experiments.
\end{abstract}
\firstfigfalse

\section{Introduction}

One of the long-sought goals in low-temperature physics is the
creation of two interpenetrating superfluids.  Early efforts were
directed at mixtures of helium isotopes.  More recently, following the
experiments with Bose-Einstein condensates (BEC) of atomic gases
\cite{Anderson_et_al,Bradley_et_al,Davis_et_al}, considerable efforts
have been made to create systems where two species of atoms condense
simultaneously.  This goal was partially achieved for two different
hyperfine spin states of $^{87}$Rb, which were condensed in the same
trap by the technique of sympathetic cooling \cite{Myatt_et_al}.
Later the dynamics of the complex relative motion of the condensates
has been studied \cite{Hall_et_al1}.  The possibility of the
measurement of the relative phase between the two condensates has also
been demonstrated \cite{Hall_et_al2}.  In these experiments the two
condensates have a substantial overlap, although they do not
completely interpenetrate each other in the stationary state.  A
similar state called ``spinor condensate'' has been observed for
sodium gas \cite{Ketterle}.

Theoretical investigation of two-component Bose systems has started
many decades ago with the construction of the phenomenological
hydrodynamic equations in the spirit of the Landau-Khalatnikov
two-fluid model for the one-component BEC \cite{Khalatnikov}.  Later,
this construction has been put onto a microscopic basis
\cite{AndreevBaskin,Mineev,Nepomnyashchii}.  Recent experiments with
alkali atoms have revived the interest in the theory of such systems.
Hartree-Fock theory has been successfully tested on the two-component
$^{87}$Rb system \cite{HFtheory}.  The stability \cite{stability},
ground-state properties \cite{groundstate} and collective excitations
\cite{excitations} have been studied theoretically by using the
Gross-Pitaevskii equations.

Many properties of two-component, or binary, BEC can be understood from 
symmetry
arguments.  Compared to one-component Bose superfluids, two-component
systems have more interesting pattern of symmetry and symmetry
breaking.  Condensation in binary Bose systems corresponds to 
the spontaneous breaking
of {\em two} (instead of one) global U(1) symmetries.  These symmetries
are related, by Noether's theorem, to the separate conservation of the
number of atoms of each of the two species.  The quantum 
state of the binary Bose
system, therefore, is characterized by two phases of the two condensates.
Correspondingly, the physics of binary BEC is also richer than of
usual one-component systems.

The effects of a symmetry are often best exposed by violating the
symmetry explicitly in a controlled fashion.
A very interesting feature, specific to systems consisting of atoms of
the same isotope in different spin states, is that it is possible to
couple two condensates by a driving electromagnetic field tuned to
the transition frequency.  In this case atoms can be interconverted
between the two spin states and the numbers of atoms of each species
are not conserved separately anymore; only the total number of atoms is
constant.  This implies that, in the presence of the coupling drive,
only one U(1) symmetry remains exact, the other one is explicitly
violated.  The preserved U(1) symmetry obviously comes from the
conservation of the total number of atoms, and corresponds to changing
the phases of the two condensates by the same amount (i.e., leaving
the relative phase unchanged).  The violated U(1) corresponds to
changing the relative phase between the two condensate.  The presence
of the coupling drive lifts the degeneracy of the ground state
with respect to the relative phase.

In this work, we show that a sufficiently small violation of the U(1)
symmetry corresponding to the relative phase leads to the existence of
a nontrivial static configuration --- a domain wall inside which the
relative phase changes by $2\pi$.  This configuration is a local
minimum of the energy.  However, the domain wall is {\em not}
topologically stable and can ``unwind'' itself.  To unwind, however, 
the system must overcome an energy barrier.  Thanks to this
fact, the rate of the spontaneous decay 
of the domain wall is exponentially suppressed.

Our paper is organized as follows.  Section \ref{sec:L} introduces the
field-theoretical description of binary BEC.  In Sec.\
\ref{sec:solution} we describe the domain wall configuration, whose
physical interpretation is given in Sec.\ \ref{sec:interpretation}.
Section \ref{sec:boundary} deals with the boundary of finite domain
walls and the related phenomenon of ``vortex confinement''.  Section
\ref{sec:conclusion} contains concluding remarks.  In Appendix
\ref{sec:stability} we find the metastability condition for the
domain wall in the particular case when the densities of the two
components are equal, and in Appendix \ref{sec:decay} two different
mechanisms for the decay of the domain wall, operating at different
temperature regimes, are considered.

\section{The Lagrangian, its symmetries and normal modes}
\label{sec:L}

In this Section, we use field theory to describe general properties of
binary BEC.  Our goal is to introduce notations and the formalism to
lay the ground for the discussion of the domain walls in the next
Section.

A binary dilute Bose system is described by a quantum field theory
of two complex scalar fields $\psi_1$ and $\psi_2$.  These
fields have the meaning of the wave functions of the two condensates.
The dynamics of these fields is governed by the following Lagrangian,
\begin{equation}
  L = i\hbar(\psi_1^\+\d_t\psi_1 + \psi_2^\+\d_t\psi_2)
        - H(\psi_1, \psi_2)\,,
  \label{L}
\end{equation}
where the Hamiltonian $H(\psi_1,\psi_2)$ has the form
\begin{eqnarray}
  H(\psi_1,\psi_2) & = &
      {\hbar^2\over2m} (|\nabla\psi_1|^2 + |\nabla\psi_2|^2)
      - \mu_1 \psi_1^\+\psi_1 - \mu_2 \psi_2^\+\psi_2 \nonumber\\
    & &  + {g_{11}\over2}|\psi_1|^4 
      + {g_{22}\over2}|\psi_2|^4
      + g_{12}|\psi_1|^2 |\psi_2|^2
      - {1\over2}\hbar\Omega(\psi_1^\+\psi_2+\psi_2^\+\psi_1)\,.
  \label{H-full}
\end{eqnarray}
In Eq.\ (\ref{L}) $\mu_{1,2}$ are the chemical potentials
%(in the absence of the coupling drive, $\Omega=0$)
of the two species,\footnote{In real experiments $\mu_1$ and $\mu_2$
are functions of coordinates.  We assume here that the trapping potentials
are sufficiently wide so that these chemical potentials can be put to
constants.}
 $g_{ij}$ is the scattering amplitude, in the zero
momentum limit, between an atom of the $i$-th species and that of the
$j$-th species, and are proportional to the scattering lengths $a_{ij}$,
\begin{equation}
g_{ij}=4\pi\hbar^2a_{ij}/m\,; 
\end{equation}
and $\Omega$ is the Rabi frequency
arising from the coupling drive.  

By varying the action $S=\int\!dt\,d\x\,L$ with respect to
$\psi_{1,2}$, the familiar Gross-Pitaevskii equations are directly
obtained:
\begin{eqnarray}
  i\hbar\d_t\psi_1 &=& -{\hbar^2\over2m}\nabla^2\psi_1 - \mu_1\psi_1 +
  (g_{11}|\psi_1|^2 + g_{12}|\psi_2|^2)\psi_1 - {1\over2}\hbar\Omega\psi_2\,;
  \nonumber \\
  i\hbar\d_t\psi_2 &=& -{\hbar^2\over2m}\nabla^2\psi_2 - \mu_2\psi_2 +
  (g_{12}|\psi_1|^2 + g_{22}|\psi_2|^2)\psi_2 - {1\over2}\hbar\Omega\psi_1\,.
  \label{GP}
\end{eqnarray}
%Eqs.\ (\ref{GP}) are Heisenberg operator equations; by taking the
%average, they become the classical equations on the condensate
%wave functions.

Let us start by finding the ground state when the coupling drive is
off, $\Omega=0$.  In the superfluid ground state, both $\psi_1$
and $\psi_2$ have nonzero expectation values.  These can be found by
minimizing the potential energy part in Eq.\ (\ref{L}) with respect to
$\psi_1$ and $\psi_2$.  This minimization procedure gives the
equations determining the densities $n_1=|\psi_1|^2$, $n_2=|\psi_2|^2$
in terms of the chemical potentials $\mu_1$ and $\mu_2$,
\begin{eqnarray}
  g_{11} n_1 + g_{12} n_2 &=& \mu_1 \,,\nonumber\\
  g_{12} n_1 + g_{22} n_2 &=& \mu_2 \,.\label{nmu}
\end{eqnarray}
More conveniently, one could view Eq.\ (\ref{nmu}) as the equations
fixing the chemical potentials for given values of the densities.

Strictly speaking, Eqs.\ (\ref{nmu}) only correspond to an {\em
extremum} of the potential energy.  For it to be a local minimum, the
quadratic form $g_{11}n_1^2+g_{22}n_2^2+2g_{12}n_1n_2$ needs to be
positive definite:
\begin{equation}
  g_{11} g_{22} - g_{12}^2 > 0 \,.
  \label{stability}
\end{equation}
In fact, Eq.\ (\ref{stability}) is the condition for the mixture of the 
two Bose
superfluids to be thermodynamically stable against segregation
\cite{Mineev,HFtheory}.  In this paper we shall assume that
(\ref{stability}) is satisfied.

In principle, the constants $g_{11}$, $g_{12}$ and $g_{22}$ can be
arbitrary.  In this paper we limit ourselves to the
regime when all three scattering lengths are close to each other,
$a_{11}\approx a_{12}\approx a_{22}$.  
In the case when the two species are
rubidium atoms in different hyperfine states, these lengths were found
experimentally to
differ by no more than a few percent.  
The assumption that $g_{11}\approx g_{12}\approx g_{22}$ also introduces
considerable technical
simplifications in our treatment.  We introduce the ``average''
scattering amplitude
\begin{equation}
  g \equiv {g_{11} + g_{22} \over 2}\,,
\end{equation}
and the deviations from the average
\begin{equation}
  \delta g \equiv -(g_{12} - g) \,, \qquad
  \delta g' \equiv g_{11}-g = -(g_{22}-g)\, ,
\end{equation}
so that $\delta g\sim\delta g'\ll g$.  The stability condition
(\ref{stability}) implies that $\delta g > 0$.  Analogously, we
introduce the average scattering lengths $a$ and the deviations
$\delta a$ and $\delta a'$.  Note that in the limit 
$g_{11}=g_{12}=g_{22}$ the Hamiltonian has an SU(2) symmetry which leads
to interesting implications \cite{Birman}.

With the Lagrangian at hand, the discussion of symmetry in the Introduction
can be made concrete.  In the absence of the coupling drive,
$\Omega=0$, the Lagrangian (\ref{L}) possesses a U(1) $\times$ U(1)
symmetry with respect to independent phase rotations of the fields,
\begin{equation}
  \psi_1 \to e^{i\alpha_1}\psi_1, \qquad
  \psi_2 \to e^{i\alpha_2}\psi_2\,.
\end{equation}
The corresponding conservation laws are those of the numbers of
particles of each species, $N_1=\int\!d^3\x\,\psi_1^\+\psi_1$ and
$N_2=\int\!d^3\x\,\psi_2^\+\psi_2$.  That $N_1$ and $N_2$ at $\Omega=0$
are conserved
separately is actually a basic assumption made when we wrote down the
Lagrangian (\ref{L}).  This assumption is not automatically satisfied:
it requires that only elastic scattering between atoms is allowed;
inelastic scattering is forbidden.  For binary Bose systems made of
rubidium atoms, this appears to be a good approximation 
%(although the
%reason for this appears to be somewhat accidental)
\cite{scattering_length}.

Once the coupling drive is turned on ($\Omega\neq0$), the Lagrangian
is invariant only under a subset of the original U(1) $\times$ U(1)
rotations; namely, those which rotate both $\psi_1$ and $\psi_2$ by
the same angle,
\begin{equation}
  \psi_1 \to e^{i\alpha} \psi_1, \qquad
  \psi_2 \to e^{i\alpha} \psi_2 \, .
\end{equation}
Therefore one of the U(1) symmetries the system enjoyed at $\Omega=0$ 
is explicitly
violated.  Applying the Goldstone theorem, we conclude that, at
$\Omega=0$, there are two
gapless excitations  and only one of these modes remain gapless at
$\Omega\neq0$.  The gapless modes at $\Omega=0$ are the phonons of the
two types of sounds. One corresponds to the ordinary density wave 
(B mode in our paper, see below), and another to the concentration wave (A
mode) in which the densities of the two
species oscillate relative to each other in such a way that the 
total density remains constant.%
\footnote{In the hydrodynamic regime these two modes would correspond to 
the first and the third sounds respectively, 
according to the established classification
 \cite{Khalatnikov}. In the zero-temperature limit the density wave
is termed Bogolyubov (zero) sound, and there is no established term for
the concentration wave.}
If we view the two components as being made of the same atoms, but
in different hyperfine levels, then the mode A can be alternatively
interpreted as a spin polarization wave, in which the density of
nuclear spin is a function of space and time.

When the coupling drive is on, only the
density wave remains gapless; the phonon of the concentration (spin
density) wave is gapped.

Let us compute the sound speeds at $\Omega=0$.  To this end we write
$\psi_1$ and $\psi_2$ as
\begin{equation}
  \psi_1 = \sqrt{n_1+\delta n_1}\, e^{i\varphi_1}, \qquad
  \psi_2 = \sqrt{n_2+\delta n_2}\, e^{i\varphi_2},
\end{equation}
and expand Eq.\ (\ref{L}) to second order of $\delta n$ and
$\nabla\varphi$ (we will see that $\delta n\sim\nabla\varphi$).  We
find
\begin{eqnarray}
  L^{(2)} &=& -\hbar(\delta n_1\d_t\varphi_1+\delta n_2\d_t\varphi_2) -
      {\hbar^2\over 2m}(n_1|\nabla\varphi_1|^2+n_2|\nabla\varphi_2|^2) 
  \nonumber \\
    & &
    -{1\over2}(g_{11}\delta n_1^2 + 2g_{12}\delta n_1\delta n_2 +
         g_{22}\delta n_2^2)
  \label{Lnphi}
\end{eqnarray}
(we have thrown away total derivatives).  The density fluctuations
$\delta n_i$ can be ``integrated out'' and replaced by the saddle point
values $\delta n_i=-\hbar\sum_j(g^{-1})_{ij}\d_t\varphi_j$.  As a result,
Eq.\ (\ref{Lnphi}) becomes
\begin{equation}
  L^{(2)} = {\hbar^2\over2}\sum_{ij}(g^{-1})_{ij}\d_t\varphi_i\d_t\varphi_j -
  {\hbar^2\over2m}(n_1|\nabla\varphi_1|^2+n_2|\nabla\varphi_2|^2)\,.
  \label{Lphi}
\end{equation}
Thus, the dispersion relations for the phonons are linear,
$\omega=uk$, and the sound speed $u$ satisfies the characteristic
equation
\begin{equation}
  \det \left( \left(g^{-1}\right)_{ij} - {1\over mu^2}\left(
    \begin{array}{cc} n_1 & 0 \\ 0 & n_2 \end{array} \right)\right)
  = 0\,.
\end{equation}
When $g_{11}\approx g_{12}\approx g_{22}$, the solutions are
\begin{equation}
  u_\A^2 = {2\delta g\over m}{n_1 n_2\over n}
\,,\qquad 
  u_\B^2 = {gn\over m} \,. %= 4\pi {\hbar^2 an\over m^2}
  %=2{\delta g\over g} {n_1 n_2\over n^2} u_1^2
  \label{soundspeed}
\end{equation}
We see that when $\delta g\ll g$ the speed of the concentration wave
(A) is much
smaller than that of the density wave (B), $u_\A\ll u_\B$. 
The system is ``stiffer'' towards density fluctuations than
towards fluctuations of concentration. That modes
A and B are indeed concentration and density fluctuations
respectively is seen from the corresponding eigenvectors.
The A mode corresponds to such fluctuations in which
\begin{equation}
  \mbox{A:}\qquad
  \left( \begin{array}{c} \varphi_1 \\ \varphi_2 \end{array}\right)
  \sim
  \left( \begin{array}{c} n_2 \\ -n_1 \end{array}\right), \qquad
  \left( \begin{array}{c} \delta n_1 \\ \delta n_2 \end{array}\right)
  \sim
  \left( \begin{array}{c} 1 \\ -1 \end{array}\right) \, ,
\end{equation}
while the B mode corresponds to 
\begin{equation}
  \mbox{B:}\qquad
  \left( \begin{array}{c} \varphi_1 \\ \varphi_2 \end{array}\right)
  \sim
  \left( \begin{array}{c} 1 \\ 1 \end{array}\right), \qquad
  \left( \begin{array}{c} \delta n_1 \\ \delta n_2 \end{array}\right)
  \sim
  \left( \begin{array}{c} n_1 \\ n_2 \end{array}\right) \, .
\end{equation}
Therefore, in the A sound $n_1$ and $n_2$ fluctuate in such a way that
the overall density remains constant ($\delta n_1+\delta n_2=0$),
while the B sound corresponds to density waves in which
$n_1/n_2$, or concentration, is unchanged ($\delta n_1/n_1=\delta n_2/n_2$).  
The Lagrangian (\ref{Lphi}), in
terms of the normal modes
\begin{eqnarray}
  \phiA \equiv \varphi_1 - \varphi_2 \quad\mbox{and}\quad%\nonumber\\
  \phiB \equiv (2/ n)(n_1\varphi_1 + n_2\varphi_2)\,,
  \label{phiAB}
\end{eqnarray}
 has the form
\begin{equation}
  L^{(2)} = {\hbar^2n\over8m} \biggl\{ 
      {4n_1n_2\over n^2}\left[
        u_\A^{-2}(\d_t\phiA)^2-(\nabla\phiA)^2\right]+
      \left[
        u_\B^{-2}(\d_t\phiB)^2-(\nabla\phiB)^2\right]
\biggr\}\,.
\end{equation}

When the coupling drive is on, $\Omega\neq0$, one should add
to Eq.\ (\ref{Lphi}) the potential energy term
$\hbar\Omega\sqrt{n_1n_2}\cos(\varphi_1-\varphi_2)$.%
\footnote{
In the Hamiltonian approach, the effect
of the coupling on the spectrum 
can be treated by diagonalizing Hamiltonian
using an exact transformation \cite{referee}.
}
  The B sound
is not affected by this term since it corresponds to such fluctuations
in which $\varphi_1=\varphi_2$.  The phonons of the A sound
acquire a gap
\begin{equation}
  \Delta = \sqrt{2\hbar\Omega\cdot \delta gn}
    \biggl( {n_1n_2\over n^2}\biggr)^{1/4}
\end{equation}
for small values of $\Omega$.

In our further discussion we shall need the formulas for the healing, 
or correlation, lengths, 
which are defined via the response of the system to a static
source coupled locally to the particle density,
$\mu_1(\x)|\psi_1|^2+\mu_2(\x)|\psi_2|^2$.  
As in the case of the sounds, there are also
two healing lengths,
\begin{eqnarray}
  \xi_\A &=& {1\over\sqrt8}{\hbar\over \sqrt{m\delta gn}}
    \biggl({n^2\over n_1n_2}\biggr)^{1/2} = 
    {1\over\sqrt2}\biggl({g\over\delta g}\biggr)^{1/2} 
    \biggl({n^2\over n_1n_2}\biggr)^{1/2}\xi_\B \, , \nonumber\\
  \xi_\B &=& {1\over2}{\hbar\over\sqrt{mgn}} = 
  {1\over\sqrt{16\pi an}} \, .
  \label{coh_lengths}
\end{eqnarray}
As seen from Eq.\ (\ref{coh_lengths}), $\xi_\A\gg\xi_\B$. 
This is because $\xi_\B$ is the correlation length of fluctuations of the
overall density, while $\xi_\A$ is the correlation length of the
relative density.

If we take the values typical for present day experiments: $n\sim
10^{14}$ cm$^{-3}$, $a\sim 50$ \AA, and $\delta g/g\sim 10^{-2}$, then
$\xi_\B\sim 0.2$ $\mu$m, and $\xi_\A\sim 3$ $\mu$m.
These lengths are smaller than the typical system size in experiments.

\section{The domain wall of relative phase}
\label{sec:solution}

The existence of the domain wall of relative phase in binary BEC with 
a coupling drive
can be shown in a rather simple way. Let us first focus only on
fluctuations of the fields on length scales much larger than the 
largest healing
length $\xi_\A$.  In this case, the amplitudes $n_1$ and $n_2$ 
of $\psi_{1,2}$ can
be regarded as ``frozen'' and the only important degrees of freedom
are the phases, i.e., $\varphi_1$ and $\varphi_2$.  The energy of
the system is a functional of $\varphi_1$ and $\varphi_2$,
\begin{equation}
  E[\varphi_1,\varphi_2] = 
  \int\! d^3\x\,\biggl[ {\hbar^2\over2m}(n_1(\nabla\varphi_1)^2
      + n_2(\nabla\varphi_2)^2) 
      - \hbar\Omega\sqrt{n_1n_2}\cos(\varphi_1-\varphi_2) \biggr]\,.
  \label{H}
\end{equation}
The potential energy term
$-\hbar\Omega\sqrt{n_1n_2}\cos(\varphi_1-\varphi_2)$ has its minimum at
$\phiA\equiv\varphi_1-\varphi_2=0$.  The configuration in which
$\phiA=0$ over the whole space is obviously the global minimum of the
total energy, and hence is the ground state.  

The domain wall solution that we shall describe is, in contrast, a {\em
local} minimum of the energy.  To find the profile of the domain wall,
we vary Eq.\ (\ref{H}) with respect to $\varphi_1$ and $\varphi_2$ and
obtain the following equations
\begin{equation}
   {\hbar^2\over m}n_1\nabla^2\varphi_1 =
  -{\hbar^2\over m}n_2\nabla^2\varphi_2 =
  \hbar\Omega\sqrt{n_1 n_2} \sin(\varphi_1-\varphi_2)\,.
  \label{walleq}
\end{equation}
%In particular, $\bfnab\cdot(n_1\bfnab\varphi_1+n_2\bfnab\varphi_2)=0$,
%which is nothing but the conservation of the total number of atoms.
For domain walls, $\varphi_1$ and $\varphi_2$ are functions of only
one coordinate, say, $z$, and at $z=\pm\infty$ both approach constant
values.  A nontrivial 
solution to Eq.\ (\ref{walleq}) satisfying these conditions
is
\begin{eqnarray}
  \varphi_1 &=& {n_2\over n}\phiA, \qquad 
  \varphi_2 = - {n_1\over n}\phiA \, , \nonumber \\
\mbox{where}\qquad
  \phiA &=& 4\arctan e^{kz},\qquad 
  k^2 = {m\Omega\over\hbar}{n\over\sqrt{n_1 n_2}} \, .
  \label{profile} 
\end{eqnarray}
The characteristic width of the
domain wall is $k^{-1}$.  The tension of the domain wall is
\begin{equation}
  \sigma = 8 {\hbar^{3/2}\Omega^{1/2}n\over m^{1/2}}
           \biggl( {n_1n_2\over n^2}\biggr)^{3/4}\,.
  \label{tension}
\end{equation}

The relative phase $\phiA$ changes from 0 to $2\pi$ as $z$ runs from
$-\infty$ to $+\infty$.  Note that $\phiA$ is defined modulo $2\pi$ so
it goes a full circle as one passes through the wall.  Therefore from the
point of view of the energy functional (\ref{H}) the domain wall
(\ref{profile}) is a topologically nontrivial configuration, which can not be
continuously deformed into the $\phiA=0$ configuration. In fact, one can
prove that the domain wall is a configuration with minimal energy
defined in Eq.\ (\ref{H}) among those where $\phiA$ changes by
$2\pi$ from $z=-\infty$ to $z=+\infty$, and hence, cannot decay away,
{\em as long as Eq.\ (\ref{H}) applies}.
The domain wall is similar to the soliton of the sine-Gordon model.
There is a small difference: the ground states on two sides of the
domain wall are {\em different}: at $z=-\infty$
$\varphi_1=\varphi_2=0$, while the ground state at $z=+\infty$ is
$\varphi_1=2\pi n_2/n$, $\varphi_2=-2\pi n_1/n$.  Since
$\varphi_{1,2}$ are defined mod $2\pi$ the latter is equivalent to
$\varphi_1=\varphi_2=2\pi n_2/n$.

In reality, Eq.\ (\ref{H}) is not the full Hamiltonian of the system:
it is only an effective description valid at length scales larger than
both healing lengths.  The full theory (\ref{H-full})
contains, besides $\varphi_1$
and $\varphi_2$, also the density fluctuations $\delta n_1$ and
$\delta n_2$.  As a consequence, in the full theory, the domain wall
(\ref{profile}) {\em can} be continuously deformed into the trivial 
configuration $\phiA=0$.  Such deformations necessarily pass
through field configurations where either $n_1$ or $n_2$ vanishes at
some points: at these points $\phiA$ is ill-defined.  Thus, the domain wall
is not truly topological and can ``unwind'',
i.e., decay away.  The fact that the ground states on the two sides of the wall
are different does not prevent the decay: it is possible to construct
a field configuration interpolating between $\varphi_1=\varphi_2=0$
and $\varphi_1=\varphi_2=2\pi n_2/n$ with arbitrarily small energy per
unit area.

Although the domain wall is not a global minimum of the energy
functional, it can still be a {\em local} minimum.  In this case, to
deform the domain wall into a ``topologically'' trivial configuration
with $\phiA=0$ one has to overcome an energy barrier.  The wall is in
this case metastable.  From our previous discussion one can conclude that,
roughly speaking, the wall is metastable when Eq.\ (\ref{H}) applies
and is unstable when Eq.\ (\ref{H}) is not applicable.  For Eq.\
(\ref{H}) to be valid the wall has to be wider than the largest healing
length, $\xi_\A$.  Since the width of the wall decreases as one
increases the Rabi frequency $\Omega$, the wall is metastable only
when $\Omega$ is smaller than some critical value $\Omega_c$.  Let us
define $\Omega_0$ as the value of the Rabi frequency at which the
width of the wall $k^{-1}$, as defined in Eq.\ (\ref{profile}), is
equal to to the longer healing length $\xi_\A$ in Eq.\
(\ref{coh_lengths}):
\begin{equation}
  \hbar\Omega_0 = 8\delta g n \biggl( {n_1n_2\over n^2}\biggr)^{3/2}\,.
  \label{Omega0}
\end{equation}
The wall is metastable when $\Omega$ is less than some critical value
$\Omega_c$ of order $\Omega_0$,
\begin{equation}
  \Omega_c \sim \Omega_0\,.
  \label{Omegac}
\end{equation}
To find the exact value of $\Omega_c$ one needs to perform a more
refined stability analysis.  We present such an analysis for the special
case $n_1=n_2$ (i.e., when the densities of the two species are equal)
in Appendix \ref{sec:stability}.  Parametrically, the result is
consistent with Eq.\ (\ref{Omegac}), and the ratio $\Omega_c/\Omega_0$
is found to be $1/3$.

Using the numerical values typical for experiments with Rb: $n\sim$ a few
$10^{14}$ cm$^{-3}$, $\delta a\sim 1$ \AA, one finds $\Omega_0\sim 100$
Hz.  Therefore to have a stable wall the Rabi
frequency needs to be smaller than about 100 Hz.  The domain wall cannot be
thinner than the longer correlation length $\xi_\A$, which was
estimated above to be a few $\mu$m.

Even when $\Omega<\Omega_c$, a metastable wall can still spontaneously
decay (burst).  Such a
decay, as we have said, requires overcoming a potential barrier.  At
sufficiently large temperature, the mechanism for the decay is thermal
activation (over the barrier).  
At zero or small temperatures, the mechanism of the decay
is quantum tunneling under the energy barrier.  The
decay through thermal activation, which is relevant for the
temperatures achieved in current experiments, is considered in Appendix 
\ref{sec:decay}, where it is
shown that the decay rate is exponentially suppressed, 
since a global (macroscopic) fluctuation
is required,
unless $\Omega$ is very close to $\Omega_c$.

\section{Physical interpretation of the domain wall}
\label{sec:interpretation}

\begin{figure}[tb]
\vbox
{%
\begin{center}
\leavevmode
  \def\epsfsize #1#2{0.5#1}
  \epsfbox{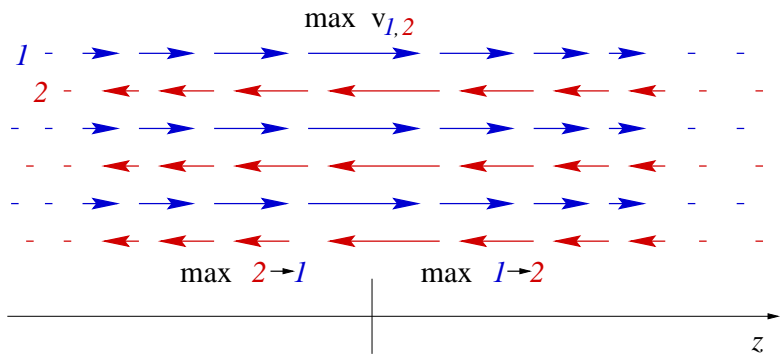}
\end{center}
  \caption{The domain wall as superfluid flows.  The two components flow in
  opposite directions perpendicular to the wall. 
  The velocities are maximal at the center
  of the wall.  The particle number of each type is not conserved due
  to interconversion which occurs inside the wall, with maximal rates
	on two sides of the center.}
\label{fig:flow}
}%
\end{figure}

The domain wall solution found in Sec.\ \ref{sec:solution} allows an
interesting physical interpretation which suggests a possible way for
their creation in experiments \cite{Friedberg}.  We first note that in a BEC
the superfluid velocity is proportional to the gradient
of the phase.  In a two-component BEC, there are two such velocities.
\begin{equation}
  \v_1 = {\hbar\over m} \bfnab \varphi_1 \, , \qquad
  \v_2 = {\hbar\over m} \bfnab \varphi_2 \, .
\end{equation}
Equation (\ref{walleq}), in particular, implies that the total
particle number current vanishes, $n_1\v_1+n_2\v_2=0$.  Individually,
however, the particle number current of each species $\j_1\equiv
n_1\v_1$ and $\j_2\equiv n_2\v_2$ are nonzero.  From Eq.\
(\ref{profile}) we find
\begin{equation}
  \v_1 = {n_2\over n}{2k\over\cosh kz}\widehat{\z} \, , \qquad
  \v_2 =-{n_1\over n}{2k\over\cosh kz}\widehat{\z}\,.
\end{equation}
Thus the domain wall is a configuration where the two components flow
in opposite directions.  The flow is illustrated in Fig.\
\ref{fig:flow}.  The velocities of the components are largest at the
center of the wall ($z =0$) and decrease as one moves toward the edge
of the wall.  The flow is concentrated on the wall; outside the wall
($|z|\gg k^{-1}$) there is essentially no flow.

Equation (\ref{walleq}) can be rewritten in terms of the currents as
\begin{equation}
  \bfnab\cdot\j_1 = -\bfnab\cdot\j_2 = \Omega\sqrt{n_1n_2}\sin\phiA\,.
  \label{conversion}
\end{equation}
For a stationary configuration as the one we are considering, Eq.\
(\ref{conversion}) means that the number of particles in each species
is not conserved.  It also implies that there is a {\em conversion} of
atoms between the two energy levels due to the coupling drive.  In the
left half of the wall $\phiA<0$ and there is a conversion of atoms of
the second type to atoms of the first type.  In the right half
$\phiA>0$ and the conversion goes the opposite way (Fig.\
\ref{fig:flow}).  The rate of
conversion is
\begin{equation}
  \Omega\sqrt{n_1n_2}\sin\phiA = -2\Omega\sqrt{n_1n_2}\,.
  {\sinh kz\over\cosh^2kz}\,.
  \label{conversionrate}
\end{equation}
As is the flow, the conversion rate is also maximal near the wall.  Far
from the wall ($|z|\gg k^{-1}$) there is essentially no conversion.
The conversion rate (\ref{conversionrate}) changes sign at $z=0$.
Since different species correspond to different energy levels of the
atom, energy is absorbed in one half of the wall and released in the other.
\footnote{
Recalling that, in current experiments, atoms 1 and 2 correspond
to different nuclear spin states of the same atom, we can
view the wall as a stationary configuration with a flow
of (nuclear) spin across $z=0$ plane accompanied by 
spin flips on both sides of it.
}

The interpretation of the domain walls given above suggests a possible
method for their creation in experiments \cite{Friedberg}.  One starts with
the coupling drive off ($\Omega=0$) and prepares a state where
the two condensates flow in opposite directions (for example, by
manipulating the traps).  In such a state the relative phase
$\phiA=\varphi_1-\varphi_2$ is a linear function of the coordinate
along the direction of motion (say, $z$).  One then slowly increases
$\Omega$.  The domain walls will be created and the centers of the
walls are located at the points where $\phiA$ was an odd multiple of
$\pi$ ($\pm\pi$, $\pm3\pi$, $\pm5\pi$, etc.) before $\Omega$ was turned
on. By changing the velocity
of the initial relative motion of the condensates and the final values
of $\Omega$ one can change the separation between the domain walls and
their width.  Such controlled creation of the domain walls, hopefully,
can be achieved in future experiments.

\section{Vortex as the boundary of the wall and vortex confinement}
\label{sec:boundary}

So far we have always considered infinite domain walls which have no
boundary.  It is also interesting, and perhaps more realistic, to
consider domain walls with a boundary.  We shall show that the domain
wall can be bounded by a vortex line.

\begin{figure}[hbt]
\vbox
{%
\begin{center}
\leavevmode
  \def\epsfsize #1#2{0.5#1}
  \epsfbox{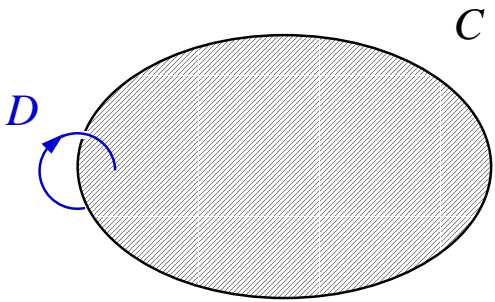}
\end{center}
  \caption{The boundary of a finite domain wall.}
\label{fig:boundary}
}%
\end{figure}

Suppose we have a finite domain wall whose boundary is a closed
contour ${\cal C}$ (Fig.\ \ref{fig:boundary}).  We shall assume that
the length of ${\cal C}$ is much larger than the width of the wall
$k^{-1}$ so one can view the domain wall as an infinitely thin
membrane stretched on ${\cal C}$ (we shall call this picture the
``thin-wall approximation'').  Let us now take another, smaller
contour which has a nontrivial linking with ${\cal C}$ (${\cal D}$ in
Fig.\ \ref{fig:boundary}).  As one goes along ${\cal D}$, one crosses
the membrane once, so the relative phase $\phiA$ changes by $2\pi$.
This is exactly what one expects from a vortex.  Therefore, ${\cal C}$
can be a vortex line.  We recall that the size of the core of the vortex
is the healing length $\xi_\A$, which is smaller or of the same order
as the width of the wall.  Therefore, in the thin wall approximation,
we have an infinitely thin membrane bounded by an infinitely thin
vortex line.    One should
note that such a bounded domain wall will tend to shrink to reduce its
energy, which comes from the wall tension and the tension of the
boundary vortex.

Conversely, for small non-zero $\Omega$, a vortex must have a domain wall
attached to it to minimize the energy due to the nontrivial phase
$\phiA$ winding around it. This means that the energy of a single
vortex per unit length is increasing linearly with the size
of the system in the transverse direction. This is in contrast to the
situation at $\Omega=0$ when the vortex tension has only a
logarithmic dependence on the size of the system.

Note that there are two types of vortices in binary BEC.
Those of the first type, which we shall call the $\varphi_1$ vortices,
have the condensate $\psi_1$ vanishing at the vortex center, while
$\psi_2$ is nonzero.  Analogously the $\varphi_2$ vortices have
$\psi_2=0$ and $\psi_1\neq0$ at their centers.  As one goes around a
$\varphi_1$ vortex, $\varphi_1$ changes by $2\pi$, while $\varphi_2$
does not change, and vice versa for a $\varphi_2$ vortex.  Thus $\phiA$
changes by either $2\pi$ or $-2\pi$ for the two types of vortices, so
the domain wall can be bounded by a vortex of either type.

% Rigorously speaking, when $\Omega\neq0$ the $\varphi_1$ and
% $\varphi_2$ vortices do not exist, since it is energetically forbidden
% to make $\phiA\neq0$ far away from the vortex center as required by
% the topology of these vortices.  Such vortices have energy per unit
% length which grow as a power of the box size (in contrast to the
% situation at $\Omega=0$ where the vortex tension has only a
% logarithmic dependence on the box size).  Therefore our statement that
% the boundary of a domain wall can be a vortex has only an approximate
% meaning, which is justified in the regime $\Omega\ll\Omega_c$.

\begin{figure}[hbt]
\vbox
{%
\begin{center}
\leavevmode
  \def\epsfsize #1#2{0.5#1}
  \epsfbox{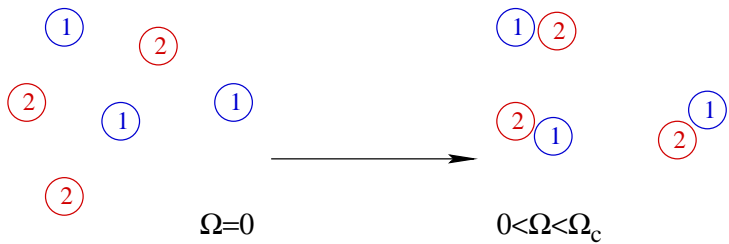}
\end{center}
  \caption{Vortex pairing (confinement).}
\label{fig:pairs}
}%
\end{figure}

In contrast to an individual vortex, a pair of $\varphi_1$ and
$\varphi_2$ vortices, placed parallel to each other, will have energy
per unit length which is only logarithmically divergent.  That is
because the $\phiA$ ``charges'' of the two vortices cancel each other,
so $\phiA$ is trivial at spatial infinity (no winding).  
The same situation occurs
for a pair of parallel vortices of the same type ($\varphi_1$ or $\varphi_2$)
with opposite winding (however, such vortex-antivortex pair can annihilate,
while a $\varphi_1\varphi_2$ pair cannot).  In a certain sense, one can
talk about the phenomenon of ``vortex confinement'': vortices exist
only in pairs.  This confinement should, in principle, be observable
experimentally, in a rotating two-component BEC, which can be already 
created in a laboratory \cite{matthews-vortex}. The vortices
are usually identified by the density depletion at their cores,
but can be also seen as dislocations of interference fringes due
to phase singularities \cite{vortex-dislocations}.
With the coupling drive off ($\Omega=0$) such a system
contains an equal number of $\varphi_1$ and $\varphi_2$ vortices which
are distributed in space with no particular correlation
between $\varphi_1$ and $\varphi_2$ vortices.  As one turns on $\Omega$,
the vortices will start to pair up and at some point the system will
become a collection of composite objects, each being a bound state of
a $\varphi_1$ and a $\varphi_2$ vortex (Fig.\ \ref{fig:pairs}).

The phenomenon of vortex confinement is very similar to that of quark
confinement in the theory of strong interaction (quantum
Chromodynamics).  Similarly to our vortices, quarks and antiquarks do
not exist as individual objects, but are confined into
composite objects ---  hadrons.  The analogy with quantum
Chromodynamics actually stretches further.  If one places a
$\varphi_1$ vortex and a $\varphi_2$ vortex at a distance larger than
$k^{-1}$, then a domain wall that connects these two vortices will be
formed.  The tension of the domain wall is the force, per unit length,
that attracts the two vortices.  The attractive force between the two
vortices is thus independent of their separation, given that the
latter is larger than the width of the domain wall.  This is analogous
to the confining force between a quark and an antiquark, which is
also constant at large distances. A confinement model which resembles
most the confinement of the vortices is the three-dimensional
compact quantum electrodynamics considered in Ref.\ \cite{polyakov}. 
In this theory the
worldlines of electrically charged particles 
are analogous to the vortices in BEC.

One can also imagine a system of two vortices which rotate
around each other so that the confining force (from the domain wall)
balances the centrifugal force.  Such a system is analogous to the
high-spin meson states in hadronic physics where a quark and an
antiquark rotate around each other.

\section{Conclusion}
\label{sec:conclusion}

We have shown that in a system of two interpenetrating BEC with a
coupling drive, there exists a domain wall solution.  The relative
phase between the two condensates changes by $2\pi$ as one travels
through the wall.  The wall solution is formally similar to the kink in
the sine-Gordon field theory, yet it is not topologically stable and
can decay. In this respect, the wall is more similar to a soap film,
which can spontaneously burst.

From the mathematical point of view, the domain walls discussed in
this paper are similar to the ones which have been studied in particle physics.
Such domain walls appear at least in two contexts: in the theory of the
hypothetical axion \cite{axions} and in high-density quark matter
\cite{SSZ}.\footnote{%Both cases are under perturbative control
Another case discussed in particle-physics literature, 
domain walls in zero-density QCD \cite{JZ}, is beyond theoretical
control.}  The similarity is that the domain wall solution arises from the
spontaneous breaking of an approximate U(1) symmetry.  In all cases
the domain wall exists only if the explicit violation of the U(1)
symmetry (determined by the value of the Rabi frequency in our case)
is small enough.  The decay of the wall in all the examples occurs via
hole nucleation.

As to the experimental realization of the domain wall, one could be
optimistic since the estimated critical value of the Rabi frequency, for
densities and scattering lengths typical for the rubidium gas in
recent experiments, is of the order 100 Hz which is not too small.  The
width of the wall, which might be as small as a few $\mu$m if the Rabi
frequency is not much smaller than critical, can be also accommodated
inside condensates of the size characteristic of present-day
experiments.  The apparent immediate obstacle is still the creation of
a system where two BEC truly interpenetrate.  In the recent
experiments with trapped atomic gas the region of overlap between the
two BEC is still small.  One can hope the technical problems of making
a genuine two-component BEC to be solved in the near future which
would enable one to study the domain walls experimentally.

\acknowledgments

The authors thank Richard Friedberg for discussions.  We thank
RIKEN, Brookhaven National Laboratory, and U.S.\ Department of Energy
[DE-AC02-98CH10886] for providing the facilities essential for the
completion of this work.  The work of DTS is supported, in part, by a
DOE OJI grant.

\appendix

\section{Wall's local stability, $\lowercase{n_1=n_2}$}
\label{sec:stability}

In the case when the densities of the two components are
equal, $n_1=n_2$, the maximal frequency  $\Omega_c$, below which
the wall configuration is still locally stable,
can be found analytically.  For illustrative purposes we shall consider
this particular case in details.

Let us recall that when the scattering lengths $a_{11}$, $a_{12}$, and
$a_{22}$ are approximately equal (assuming $a_{11}a_{22}-a_{12}^2>0$)
there are two healing lengths $\xi_\A$ and $\xi_\B$.  The healing
length related to fluctuations of the overall density $\xi_\B$ is much
smaller than the one related to fluctuations of the relative density
$\xi_\A$: $\xi_\B\ll\xi_\A$.  We shall be interested in the case when the
width of the wall $k^{-1}$ is much larger than $\xi_\B$ (but we shall
not presume any relation between $k^{-1}$ and $\xi_\A$). In this case,
the total density $n=|\psi_1|^2+|\psi_2|^2$ can be considered as frozen,
and the system can be described in terms of three variables: $\theta$,
$\varphi_1$ and $\varphi_2$,
\begin{equation}
  \left(\begin{array}{c} \psi_1\\ \psi_2 \end{array}\right) =
  \sqrt{n}\left(\begin{array}{c} 
    \displaystyle{\cos(\theta/2) e^{i\varphi_1}}\\ 
    \displaystyle{\sin(\theta/2) e^{i\varphi_2}} 
  \end{array}\right)\,,
\end{equation}
where $\theta$ runs from $0$ and $\pi$.  In terms of these three
variables, the Lagrangian has the form
\begin{equation}
  L = -\hbar n[\cos^2(\theta/2)\,\d_t\varphi_1+
      \sin^2(\theta/2)\,\d_t\varphi_2]-H\,,
  \label{Lthetaphi}
\end{equation}
where $H$ is the following Hamiltonian
\begin{eqnarray}
  H(\theta,\varphi_1,\varphi_2) &=& 
    {\hbar^2n\over2m}\biggl[{1\over4}(\nabla\theta)^2+
      \cos^2{\theta\over2}(\nabla\varphi_1)^2
    + \sin^2{\theta\over2}(\nabla\varphi_2)^2 \biggr]
    + {1\over2}(\delta\mu n - \delta g' n^2)\cos\theta \nonumber\\ 
    && - {1\over4}\delta gn^2\sin^2\theta
    - {1\over2}\hbar\Omega n\sin\theta\cos(\varphi_1-\varphi_2)\,.
  \label{Hthetaphi}
\end{eqnarray}
In Eq.\ (\ref{Hthetaphi}) $\delta\mu=\mu_1-\mu_2$.  The ground state
is found by minimizing the potential term in Eq.\ (\ref{Hthetaphi}).
From here on we consider the special case when in the ground state the
density of atoms of the two species are equal, $n_1=n_2=n/2$ (or
equivalently $\theta=\pi/2$.)  This requires $\delta\mu=\delta g'n$.
In this case $H$ possesses a discrete symmetry with respect to replacing
$\theta\to\pi-\theta$, $\varphi_1\leftrightarrow\varphi_2$.  This
symmetry is what makes it possible to find $\Omega_c$ analytically.

It is more convenient to use, instead of $\varphi_{1,2}$, the normal
modes $\varphi_{\rm A,B}$ defined in Eqs.\ (\ref{phiAB}), which in the
case $n_1=n_2=n/2$ have the form
\begin{eqnarray}
  \phiA &=& \varphi_1 - \varphi_2 \, ,\nonumber \\
  \phiB &=& \varphi_1 + \varphi_2 \, .
\end{eqnarray}
In terms of these variables the Hamiltonian becomes
\begin{eqnarray}
  H &=& {\hbar^2n\over8m} \left[ (\nabla\theta)^2 
      + (\nabla\phiA)^2 + (\nabla\phiB)^2
      + 2 \cos\theta \bfnab\phiA\cdot\bfnab\phiB\right]
  \nonumber \\
  & &  - {1\over4}\delta gn^2\sin^2\theta
    - {1\over2}\hbar\Omega n\sin\theta\cos\phiA\,.
  \label{HphiAB}
\end{eqnarray}
In order to find the domain wall configuration, we need to extremize
the energy with respect to $\theta$ and $\varphi_{{\rm A},{\rm B}}$.
Varying with respect to $\phiB$, one finds
\begin{equation}
  \bfnab\cdot(\bfnab\phiB+\cos\theta\bfnab\phiA) = 0\,.
  \label{phiBthetaphiA}
\end{equation}
This equation determines $\phiB$ for given $\theta$ and $\phiA$.  The
task of solving Eq.\ (\ref{phiBthetaphiA}) becomes much simpler if one
assumes that all variables depend only one coordinate $z$.  In this
case 
\begin{equation}
  \d_z\phiB + \cos\theta\d_z\phiA = 0 \, ,
\end{equation}
which can be trivially solved:
$\phiB=\int^z\!dz\,\cos\theta\,\d_z\phiA$.  After eliminating $\phiB$,
the energy functional one has to minimize is
\begin{equation}
  H = {\hbar^2n\over8m} \left[ (\nabla\theta)^2 
      + \sin^2 \theta (\nabla\phiA)^2 \right] 
      -{1\over4}\delta gn^2\sin^2\theta
    - {1\over2}\hbar\Omega n\sin\theta\cos\phiA\,.
  \label{HphiA}
\end{equation}
It is easy to check that the following configuration is always a local
extremum of Eq.\ (\ref{HphiA}):
\begin{eqnarray}
  \bar\theta(z) &=& {\pi\over2} \,, \label{extremum1}\\
  \bar\phiA(z) &=& 4\arctan e^{kz}, \qquad k^2 = {2m\Omega\over\hbar}\,.
  \label{extremum}
\end{eqnarray}
Equation (\ref{extremum1}) can be guessed from the symmetry of Eq.\
(\ref{HphiA}) under $\theta\to\pi-\theta$.  To see if the domain wall
solution is a local minimum, one needs to expand $H$ in the vicinity
of (\ref{extremum1},\ref{extremum}).  One writes
\begin{eqnarray}
  \theta(z) &=& \bar\theta(z) + \tilde\theta(z)\, , \nonumber\\
  \phiA(z) &=& \bar\phiA(z) + \tilde\phiA(z) \, .
\end{eqnarray}
To the second order in $\tilde\theta$ and $\tilde\phiA$ the
Hamiltonian (\ref{HphiA}) is
\begin{equation}
  H^{(2)} = {\hbar^2n\over8m} \biggl[ (\nabla\tilde\theta)^2 + \biggl(
    {2\delta g nm\over\hbar^2} + k^2\cos\bar\phiA - (\nabla\bar\phiA)^2
    \biggr)\tilde\theta^2\biggr]
      + {\hbar^2n\over8m} \left[(\nabla\tilde\phiA)^2 + 
        k^2\cos\bar\phiA\cdot\tilde\phiA^2\right]\,.
  \label{H2ndorder}
\end{equation}
One has to find the eigenmodes of Eq.\ (\ref{H2ndorder}): if there are
no negative modes then the domain wall is a local minimum of the
energy; if there exist a negative mode then the domain wall is
unstable.  The second, $\theta$-independent, 
term in Eq.\ (\ref{H2ndorder}) does not have
negative modes (it has only one zero mode corresponding to the
translation of the wall along the $z$ direction) and does not lead to
instability, and hence can be ignored.  Taking into account the
explicit solution $\bar\phiA=4\arctan e^{kz}$, the first term in Eq.\
(\ref{H2ndorder}) is
\begin{equation}
  {\hbar^2 n\over 8m}\biggl[ (\nabla\tilde\theta)^2 +
      \biggl( {2\delta g nm\over\hbar^2} + k^2 - {6k^2\over\cosh^2kz}
      \biggr) \tilde\theta^2 \biggr]\,.
\end{equation}
The well-known operator
\begin{equation}
  -\nabla^2 - {6k^2\over\cosh^2kz}
\end{equation}
has the lowest eigenvalue equal to $-4k^2$, corresponding to the
eigenfunction $\cosh^{-2}kz$, which implies that $H$ does not have a
negative mode if $\Omega<\Omega_c$, where
\begin{equation}
  \Omega_c = {1\over3} {\delta g n\over \hbar}\,.
  \label{Omegac_exact}
\end{equation} 
When $\Omega>\Omega_c$, the configuration (\ref{extremum}) is not a
local minimum of the energy functional: the domain wall does not
exist.  The value (\ref{Omegac_exact}) is of the same order as
$\Omega_0$ in Eq.\ (\ref{Omega0}): $\Omega_c={1\over3}\Omega_0$.  The
numerical value ${1\over3}$ is specific for the case $n_1=n_2$; if
$n_1\neq n_2$ then $\Omega_c/\Omega_0$ is, in general, different.

\section{Decay of the domain wall}
%\section{Decay of the domain wall}
\label{sec:decay}

As we have seen above, as long as $\Omega<\Omega_c$, the wall
minimizes the energy of the system with respect to small {\em local}
variations of the condensates. However, the global minimum of the
energy is achieved when the phases of the condensates are constant in
space. Since, as discussed above, the wall configuration can be
continuously deformed into the global minimum energy configuration,
i.e., the wall can decay.  Here we shall estimate the lifetime of the wall
due to such a decay.

Since the wall minimizes the energy locally, such a deformation
necessarily goes through a potential barrier.  Thus the decay can only
occur by a quantum tunneling through this barrier or, at finite
temperature, by a thermal fluctuation over the barrier.  
In both cases, the decay rate is exponentially suppressed: in
the first case by the WKB factor $e^{-S/\hbar}$, and in the second
case by the Boltzmann factor $e^{-E_c/k_BT}$, where $E_c$ is the height of
the barrier.  The first formula applies at sufficiently low
temperature, while the second one applies at higher temperatures. 
A crude estimate (see below)
suggests that the corresponding crossover temperature 
%the crossover temperature between the
%quantum-mechanical rate and the thermal rate 
is quite small
(nanokelvins for parameters typical for present-day experiments),
and is lower than temperatures achieved in present-day experiments.
Thus we shall limit our discussion to the decay by
thermal fluctuation. This case is also simpler theoretically.
The quantum mechanical decay, relevant for very low temperatures,
 will be left beyond the scope of this paper.

%\subsection{Decay by a thermal fluctuation}
%\label{sec:thermaldecay}

Assuming the temperature is much smaller than the critical temperature at
which one of the condensates melts (so that most atoms are still in
the condensates),
the rate of thermal activation across the barrier is $\Gamma\sim
e^{-E_c/k_BT}$, where $E_c$ is the height of the barrier at zero temperature.  
Since we are
dealing with an infinite-dimensional configuration space, $E_c$ should
be understood as the energy at the lowest point of the barrier.  This
point is a saddle point (the energy has a single negative curvature 
direction).

Some information about the exponent $E_c/k_B T$ can be obtained by a
simple scaling argument without a detailed calculation.  
In terms of the dimensionless variables
$\widetilde\x$
defined as
\begin{equation}
  \x = {\hbar\over\sqrt{m\delta g n}}\, \widetilde \x \,,
\end{equation}
the energy becomes
\begin{eqnarray}
  E[\theta,\varphi_1,\varphi_2] 
  = {\hbar^3n^{1/2}\over m^{3/2}\delta g^{1/2}}\int\!d^3\widetilde \x\, 
     & & \biggl\{ {1\over2} \biggl[ {1\over4}(\tilde\nabla\theta)^2 +
     \cos^2{\theta\over2}(\tilde\nabla\varphi_1)^2 + 
     \sin^2{\theta\over2}(\tilde\nabla\varphi_2)^2\biggr] \nonumber\\
     & & -{1\over4} (\sin^2\theta+2\cos\theta_0\cos\theta)
     -{1\over2} {\hbar\Omega\over\delta gn} 
     \sin\theta\cos(\varphi_1-\varphi_2)\biggr\}\,,
  \label{Hrescale}
\end{eqnarray}
where $\theta_0$ determines the relative condensate densities
in the ground state
at $\Omega=0$: $\cos\theta_0 = (n_1-n_2)/n$. Consider the dependence
on $\Omega$ at a given $\theta_0$. The energy functional Eq.\
(\ref{Hrescale}) is equal to a dimensionful constant times
a dimensionless functional which depends on $\Omega$ only 
via the ratio
$\Omega/\Omega_0$, where $\Omega_0=(\delta
gn/\hbar)\sin^{3/2}\theta_0$, same as in (\ref{Omega0}). Thus 
the saddle point of the energy is also a function of this
dimensionless ratio:
\begin{equation}
  E_c = {\hbar^3 n^{1/2}\over m^{3/2}\delta g^{1/2}} 
      F\biggl({\Omega\over\Omega_0}\biggr)\,.
  \label{Escaling}
\end{equation}
If, moreover, we define a temperature $T_0$ as 
\begin{equation}
  k_BT_0 = {\hbar^2 n^{2/3}\over m}\,,
\end{equation}
which is of the same order as the critical temperature of
the Bose-Einstein condensation, then the decay rate for a given
$n_1/n_2$ (or $\theta_0$) can be
written as
\begin{equation}
  \Gamma \sim e^{-E_c/k_B T}
  = \exp\biggl[-{\hbar\over m^{1/2}\delta g^{1/2} n^{1/6}}
  {T_0\over T} F\biggl({\Omega\over\Omega_0}\biggr)\biggr]
  = \exp\biggl[-{1\over\sqrt{4\pi \delta an^{1/3}}}
  {T_0\over T} F\biggl({\Omega\over\Omega_0}\biggr)\biggr]\,.
  \label{decayscaling}
\end{equation}
The form of the function $F(\Omega/\Omega_0)$ cannot be found from
scaling arguments alone.  However, when $\Omega\sim\Omega_0$ one can
expect $F(\Omega/\Omega_0)\sim1$, and then the exponent $E_c/k_BT$ 
is large, since
$\delta an^{1/3}\ll an^{1/3}\ll1$, and $T\ll T_0$.

\begin{figure}
\vbox
{%
\begin{center}
\leavevmode
  \def\epsfsize #1#2{0.5#1}
  \epsfbox{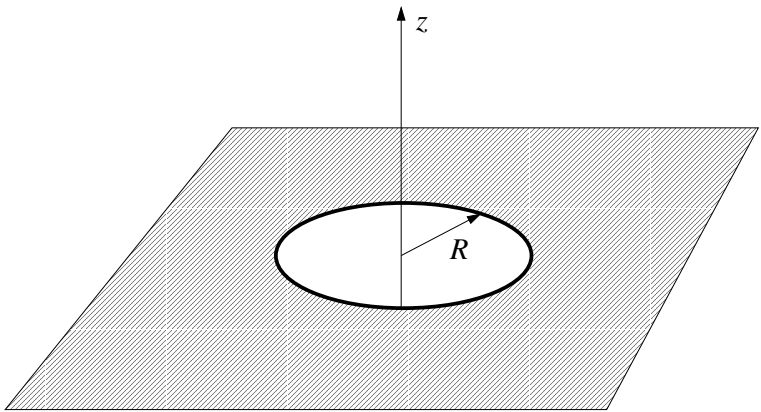}
\end{center}
  \caption{A hole in a domain wall.}
\label{fig:hole}
}%
\end{figure}

The function $F(\Omega/\Omega_0)$ can be computed in the
regime $\Omega\ll\Omega_0$, where the saddle point configuration can be
found in the ``thin-wall'' approximation.  In this approximation field
configurations are described at length scales much larger than the
width of the domain wall.  From this point of view the domain wall is
an infinitely thin membrane.  The saddle point configuration is a
membrane with a round hole in it (Fig.\ \ref{fig:hole}). The radius
of the hole $R$ must be much larger than the width of the
wall for the thin wall approximation to be valid.  As discussed in
Sec.\ \ref{sec:boundary}, the rim of the hole must be a vortex, since
it is the boundary of the domain wall.

There are two contributions to the energy difference $E_c$ between the
saddle point and the domain wall configurations.  
One contribution is negative
and comes from the hole (since the hole is the absence of the wall).
Another contribution is positive and comes from the rim.  
Therefore,
\begin{equation}
  E = 2\pi R\nu -  \pi R^2 \sigma\,,
  \label{Enusigma}
\end{equation}
where $\nu$ is the energy per unit length of the vortex (the vortex
tension) and $\sigma$ is the domain wall tension.  The energy
(\ref{Enusigma}) has a maximum when the radius of the hole is
\begin{equation}
R_c\equiv\nu/\sigma\,.
\end{equation}
This is
the radius of the critical hole.  Indeed, if a hole with a radius
$R>R_c$ is nucleated, then it will expand and eventually eat up the
whole wall.  If, in contrast, the radius of the nucleated hole is less
than critical, then the hole will shrink and disappear.

%Therefore, for either kind,
%not only $\phi_\A$ changes by $2\pi$ on a closed path threaded by the
%vortex, but also $\phi_\B$ changes by $2\pi$. This means that the
%vortex eating up the wall leaves behind a configuration in which
%$\phi_\A$ is constant while $\phi_\B$ changes by $2\pi$ along an axis
%in the $z$ direction. Because potential energy depends on
%$\phi_\B$ only through the gradient, such a configuration has 
%the energy per unit area inversely proportional to $L_z$, the size of
%the system in the $z$ direction. We shall ignore this energy compared
%to the energy of the wall.

Substituting $R=R_c$ in Eq.\ (\ref{Enusigma}), we find the height of
the energy barrier
\begin{equation}
  E_c = {\pi\nu^2\over\sigma} \,.
  \label{barrier}
\end{equation}
As there are two types of vortices, in Eq.\ (\ref{barrier}) $\nu$
refers to the vortex with the smaller tension.  The tension of a
straight vortex is logarithmically divergent: 
\begin{equation}
  \nu_i = {\hbar^2 n_i\over 2 m} \ 2\pi\ln{R\over\xi_\A}\,.
\label{nu_i}
\end{equation}
where the index $i=1,2$ refers to the two types of vortices, and $R$
is the long-distance cutoff.  The role of $R$ is played by either the
size of the critical hole $R_c$ or the width of the wall $k^{-1}$.  We
shall see at the end of this Section 
that the two lengths differ only by a logarithm,
which does not affect Eq.\ (\ref{nu_i}).  Therefore, to logarithmic
accuracy, the vortex tension is
\begin{equation}
  \nu_i = {\pi\hbar^2n_i\over m}\ln{1\over k\xi_\A}\,.
  \label{nui}
\end{equation}
The argument of the logarithm is
$(k\xi_\A)^{-1}=(\Omega_0/\Omega)^{1/2}$, and is large when
$\Omega\ll\Omega_0$, which justifies the use of the logarithmic
approximation.

Without loss of generality, we can assume that $n_1\le n_2$.  Then the
vortex of the first type has the smallest $\nu$ (\ref{nu_i}).  
Substituting Eqs.\
(\ref{nui}) and (\ref{tension}) into Eq.\ (\ref{barrier}), one finds
that the barrier height has the form of Eq.\ (\ref{Escaling}), where
\begin{equation}
  F\biggl({\Omega\over\Omega_0}\biggr) = {\pi^3\over64\sqrt2}
  {nn_1^{1/2}\over n_2^{3/2}}\biggl({\Omega_0\over\Omega}\biggr)^{1/2}
  \ln^2{\Omega_0\over\Omega}\,.
\end{equation}
The decay rate (\ref{decayscaling}) can be rewritten in the following
form
\begin{equation}
  \Gamma_{\rm thermal} \sim 
  \exp\Biggl\{-{\pi^2\over128\sqrt2}\left[\zeta\left({3\over2}\right)\right]^{2/3}
  \biggl({n\over n_2}\biggr)^{3/2}
  {1\over\sqrt{4\pi\delta an_1^{1/3}}}{T_{c1}\over T}
  \biggl({\Omega_0\over\Omega}\biggr)^{1/2}
  \ln^2{\Omega_0\over\Omega}\Biggr\}\,,
  \label{gamma}
\end{equation}
where we introduced
\begin{equation}
k_BT_{c1} = {2\pi\hbar^2\over m}\left(n_1\over\zeta(3/2)\right)^{2/3}
\end{equation}
--- the critical temperature for the smallest of the two
condensates ($n_1$ by our choice). The rate (\ref{gamma}) is exponentially
suppressed when
$\Omega\ll\Omega_0$ and $T\ll T_{c1}$.

To check the consistency of our assumptions, we note that the radius
of the critical hole is
\begin{equation}
  R_c = {\nu_1\over\sigma} 
  = {\pi\over16}\ \sqrt{\hbar\over m\Omega}
   {n\over n_2} \biggl({n_1n_2\over n^2}\biggr)^{1/4} 
  \ln{\Omega_0\over\Omega} \, .
\label{rc}
\end{equation}
Comparing with the width of the wall in Eq.\ (\ref{profile}), we find
that $R_c$ is larger than $k^{-1}$ by a factor $\ln(\Omega_0/\Omega)$,
which is assumed to be parametrically large.  Thus, the use of the
thin wall approximation is justified.

To make a good estimate of the crossover temperature, below
which crossover proceeds via quantum tunneling, rather than
via thermal activation, one needs to estimate the action on the
tunneling trajectory and compare it to the exponent in the
thermal activation rate $E_c/k_B T$. 
For a very crude estimate we can take $S\sim E_c R_c/u_A$.
The exponents $S/\hbar$  and  $E_c/k_B T$ become comparable
at a temperature of order few nanokelvin, given typical
parameters of present day experiments.

\end{document}